\documentclass[useAMS,usenatbib]{mn2e}

\usepackage{graphicx}
\graphicspath{%
    {converted_graphics/}
    {C:/Users/Mitch/Documents/W/MSS/Blandford09/converted_graphics/}
    {/}
}
\def\be{\begin{equation}}
\def\ee{\end{equation}}

\def\etal{{\rm et al.}}

\def\gtorder{\mathrel{\raise.3ex\hbox{$>$}\mkern-14mu
     \lower0.6ex\hbox{$\sim$}}}
\def\ltorder{\mathrel{\raise.3ex\hbox{$<$}\mkern-14mu
     \lower0.6ex\hbox{$\sim$}}}

\title{Radiatively Inefficient Accretion: Breezes, Winds and Hyperaccretion}
\author[Mitchell C. Begelman]
{Mitchell C. Begelman$^{1,2}$\thanks{E-mail: mitch@jila.colorado.edu}
\\
          $^1$JILA, University of Colorado and National Institute of Standards and Technology, Boulder, CO 80309-0440, USA\\
	  $^2$Department of Astrophysical and Planetary Sciences, University of Colorado, Boulder, CO 80309-0391, USA}

\begin{document}
\maketitle
\begin{abstract}
We reformulate the adiabatic inflow-outflow (ADIOS) model for radiatively inefficient accretion flows, treating the inflow and outflow zones on an equal footing.  For purely adiabatic flows (i.e., with no radiative losses), we show that the mass flux in each zone must satisfy $\dot M \propto R^n$ with $n=1$, in contrast to previous work in which $0<n< 1$ is a free parameter but in rough agreement with numerical simulations.  We also demonstrate that the resulting two-zone ADIOS models are not dynamically self-consistent without the introduction of an energy source close in to the central regions of the flow; we identify this with the energy liberated by accretion.  We explore the parameter space of non-radiative models and show that both powerful winds and gentle breezes are possible. When small radiative losses (with fixed efficiency) are included, any centrally injected energy flux is radiated away and the system reverts to a power-law behavior with $n \la 1$, where $n$ falls in a small range determined by the fractional level of radiative losses. We also present an ADIOS model for hypercritical (super-Eddington) disk accretion, in which the radiative losses are closely related to the flow geometry.  We suggest that hyperaccretion can lead to either winds or breezes.    
\end{abstract}

\section{Introduction}
\label{sec:int}
Radiatively inefficient accretion onto black holes can occur in the limits of both low and high accretion rates. Although black holes can hypothetically absorb the gravitational energy liberated during accretion (Begelman 1979), in the realistic case of a rotating accretion flow the accretion tends to be strongly inhibited, with much less gas reaching the black hole than is supplied at large distances.  The reason for this is that the outward transport of angular momentum also involves the transport of energy, which causes the gas far from the black hole to become unbound (Narayan \& Yi 1994, 1995).   

There is ample observational evidence that black holes can be ``fussy eaters."  The Galactic Center black hole appears to be accreting at a rate orders of magnitude below the Bondi rate (Melia \& Falcke 2001, and references therein), as are the black holes in some elliptical galaxies with weak X-ray sources (Di Matteo et al.~2000; Mushotzky et al.~2000; Baganoff et al.~2003).  These systems are believed to be radiatively inefficient because their densities are low enough to inhibit cooling.  At the other extreme, cooling is suppressed in hypercritical (super-Eddington) accretion flows because radiation is trapped by the large optical depth.  Hyperaccreting sources such as the Galactic binary SS433 (King, Taam \& Begelman 2000; Begelman, King \& Pringle 2006), Galactic microquasars during outburst (Mirabel \& Rodriguez 1999), and possibly ultraluminous X-ray sources (King 2008; Gladstone, Roberts \& Done 2009) all appear to produce powerful outflows.   Numerical simulations of nonradiative accretion onto black holes also show this inhibition, either through the development of a zonal structure with inflowing gas near the rotational equator, sandwiched between outflowing zones at high latitudes, or through convective motions internal to the flow (e.g., Stone, Pringle \& Begelman 1999; Igumenschev \& Abramowicz 1999, 2000; Stone \& Pringle 2001; Hawley \& Balbus 2002; McKinney \& Gammie 2002; Igumenshchev, Narayan \& Abramowicz 2003; De Villiers, Hawley \& Krolik 2003; Yuan \& Bu 2010).

Blandford \& Begelman (1999, 2004; hereafter BB99 and BB04) proposed a simple model for radiatively inefficient accretion flows, dubbed the ``adiabatic inflow-outflow solution" (ADIOS).  In their steady state, self-similar formulation, there are inflowing and outflowing zones with equal and opposite mass fluxes that vary with radius according to $\dot M \propto R^n$ with $0\leq n < 1$.  In addition, there is presumably a small constant mass flux that reaches the black hole and is accreted --- but this need not be modeled in detail.  In the one-dimensional, one-zone models developed in BB99 (and presented using a different parameterization in BB04), the vertically averaged conservation laws and dynamical constraints --- applied only to the inflowing zone --- impose well-defined relationships among the parameters describing the model, such as the specific angular momentum of the gas, its effective binding energy (characterized by the Bernoulli function), and the mass flux index $n$.  Interestingly, $n$ is not determined by these relations, but can take on any value between 0 and 1. 

In addition to analyzing some of the two-dimensional characteristics of the inflow zone (and showing that the 1D models provide an excellent first approximation), BB04 considered the outflow zone in some detail.  In particular, they showed that outflowing streamlines can be self-consistently patched onto the inflow zone at each radius.  By assuming that the streamlines are adiabatic, laminar, and inviscid, they produced a family of self-similar wind models to represent the outflow zone.

While the BB04 models provide a ``proof of concept" for ADIOS, they do not necessarily provide an accurate depiction of even the qualitative features of a realistic zonal flow.  In particular, the assumption of a wind with adiabatic, inviscid streamlines leads to very different kinds of flows in the inflow and outflow zones.  While the inflow models depict a highly turbulent, well-mixed flow, the outflow models consist of a well-ordered flow pattern, with streamlines at higher latitudes reflecting conditions much closer to the black hole than streamlines close to top of the inflow zone. For example, since the terminal velocity along each streamline scales with the escape speed at its footpoint, the outflow models of BB04 exhibit large shears in the vertical direction, with the speed approaching ``infinity" near the evacuated cone that defines the wind's centrifugal barrier (see, e.g., Fig.~6 of BB04).

Numerical simulations that show the generic characteristics of an ADIOS (e.g., Stone et al.~1999; Hawley \& Balbus 2002) often seem to exhibit much messier outflows than the BB04 models.  The outflow region appears to be as chaotic as the inflow, with the typical speeds at each radius reflecting the local Keplerian velocity. Moreover, in adiabatic simulations one does not see a range of values for $n$, but rather (where this is measurable) a tendency for the effective value of $n$ to be close to 1.  

In this paper, we present a new formulation of ADIOS in which the inflow and outflow zones are treated on an equal footing.  Given the success of the 1D inflow model as demonstrated in BB04, we adopt a one-dimensional model for each zone and mainly consider self-similar flow.  We show that the introduction of a relatively simple two-zone model has a dramatic effect on the character of the solutions: instead of obtaining solutions for any value of $n$ in the range $0\leq n <1$, we find that non-radiative solutions must have $n=1$.  Such flows are regulated by a conserved, outward flux of energy through the flow --- ignored in the BB99 and BB04 models --- and in fact require the presence of such a flux for self-consistency.  The source of this flux must be the energy liberated by the residual accretion onto the black hole.  The inflows associated with these solutions can range from marginally bound to nearly Keplerian, depending on how large the net energy flux is and how it is partitioned between the inflow and outflow zones. Outflow solutions come in two varieties: powerful winds with speeds of order the Keplerian speed and gentle, subsonic ``breezes" driven by viscous stresses.  

Flows with self-similar radiative losses have mass flux indices $n < 1$ that depend on the rate of energy loss.  These flows also depend on the existence of a central energy flux, but the net flux in this case is gradually eroded by radiative losses. Taking the limit as the radiative loss rate goes to zero, we find that these solutions also approach $n=1$.  The inflows in these cases are always relatively tightly bound.  

Finally, we present two-zone ADIOS models for hyperaccretion, i.e., super-Eddington accretion, in which the radiative losses are tied closely to the geometrical and dynamical structure of the flow.  The special form of the energy loss rate destroys self-similarity on large scales, but at sufficiently small radii we can model the radiative effects as small corrections to nonradiative flows.  By extrapolating these effects to scales where they become nonlinear, we develop an approximate picture for the flow patterns, luminosities and accretion rates of hyperaccreting systems.  We suggest that there may be two modes of hyperaccretion, corresponding to the wind and breeze solutions mentioned above.  
  
The plan of the paper is as follows.  In Section 2 we review the one-zone ADIOS model of BB99 and BB04, introduce the equations for the two-zone model, and show that two-zone models without a net energy flux are not self-consistent. In Section 3, we explore the parameter space for perfectly non-radiative two-zone models and in Section 4 we consider the effects of (self-similar) radiative losses.  In section 5 we examine the constraints associated with dissipative heating.  We present our model for hyperaccretion in section 6.  Finally, we discuss our results and summarize our conclusions in Section 7.

\section{One- and Two-zone ADIOS Models}
\label{sec:one}

\subsection{Review of the one-zone ADIOS model}

We first review the ``one-zone" ADIOS model proposed by BB99 and BB04, referring readers to Section 2.1 of BB04 for further details.  Adopting the conventions that the local accretion rate $\dot M$ and outward viscous couple $G \equiv \dot M {\cal G}$ are positive (where ${\cal G}$ is used in BB04 to represent the viscous couple per unit mass), we obtain the following equations expressing the conservation of mass, angular momentum and energy in a steady state:
\be
\label{angmom1}
{d\over dr}(G - \dot M L) = - (1 + \eta) L {d\dot M\over dr} 
\ee
\be 
\label{energy1}
{d\over dr}(G \Omega - \dot M B) =  (\beta -1) B {d\dot M\over dr},
\ee
where $L$ is the specific angular momentum, $\Omega = L/R^2$ is the angular speed, and 
\be 
\label{Bernouilli}
B={L^2\over2R^2}-{1\over R} + H
\ee 
is the Bernoulli function when ${\rm G}M = 1$ (where we use a Roman typeface for the gravitational constant to avoid confusion with the viscous couple) and $H$ is the specific enthalpy ($= \gamma P/(\gamma - 1) \rho$ for an ideal gas with adiabatic index $\gamma$).  The terms on the right-hand side of equations (\ref{angmom1}) and (\ref{energy1}) represent the rates at which angular momentum and energy are lost to the outflow, per unit radius.  For power-law solutions with $\dot M \propto R^n$, $L \propto R^{1/2}$, $B\propto R^{-1}$, these forms for the angular momentum and energy loss are fully general with the appropriate choice of the constants $\eta$ and $\beta$.

In addition to the conservation laws, the inflow is subject to a dynamical equilibrium condition based on the radial component of the momentum equation.  Given the self-similar scalings for pressure and density, $P\propto R^{n-5/2}$ and $\rho\propto R^{n-3/2}$ (BB99), and assuming that the inflow is very subsonic, we obtain 
\begin{equation}
\label{omr}
{L^2 \over R^2}-{1\over R}+(5/2-n){P\over\rho}=0.
\end{equation}
Eliminating $P/\rho$ in favor of $B$, one finally obtains a three-parameter family of ADIOS models.  The independent parameters may be taken to be $LR^{-1/2}$, $BR$ and $GR^{-3/2}$, or $n$, $\eta$ and $\beta$, or any combination of these. 

BB04 show that solutions can be found for any value of $n$ in the interval $0\leq n < 1$.  In this interval, the constants of integration for the conservation equations may be neglected and one obtains simple relations among the various quantities.  In particular, one can show that the Bernoulli function satisfies
\be 
\label{BG}
B = {GL\over \dot M R^2} {1-n \over 1 - n\beta}  .
\ee   
Since all quantities on the right-hand side of eq.~(\ref{BG}) are positive with the possible exception of $1-n\beta$, we must have $\beta > 1/n$ in order for $B$ to be negative and the inflow to be bound.  But one can also show that $\beta \rightarrow 1$ as $n\rightarrow 1$.  The sign of $B$ therefore appears to be undefined in this limit.  Indeed, if one integrates the energy conservation equation in the limit $n\rightarrow 1$, one finds that the left-hand side must integrate to a constant, whereas the right-hand side diverges logarithmically if $\beta \neq 0$.  We conclude that the constant of integration --- which physically represents a conserved radial energy flux through the disk --- cannot be neglected in this limit. 
 
\subsection{Two-Zone Adiabatic Inflow-Outflow Models}

Let us now generalize the one-dimensional ADIOS formulation to include the outflow as well as the inflow.  In BB04, the outflow was treated in a fundamentally different way from the inflow.  By patching an adiabatic, inviscid streamline onto the inflow at every radius, BB04 showed how the outflowing gas could carry away the required amounts of angular momentum and energy.  However, the resulting flow patterns exhibit enormous shear in the poloidal ($\theta$) direction and large angular gradients of temperature and density.  This is because each streamline reflects the conditions prevailing at the radius where it was launched: streamlines launched near a given radius will lie close to the inflow-outflow boundary and have speeds close to the escape speed at that radius, whereas streamlines closer to the axis will have come from smaller radii and have correspondingly larger speeds and higher specific entropies. 

Simulations, however, suggest that this view of independent, inviscid streamlines may not be accurate.  They suggest that the outflow is as chaotic as the inflow and is much better described as being ``well-mixed."  This suggests that it might be better to treat the mean properties of the outflow at each radius, rather than trying to keep the properties of the individual streamlines separate.  In this spirit, we treat the outflow using the same parametrization as the inflow.  Denoting the inflowing (outflowing) zones by subscripts 1 (2), and keeping the sign conventions from Section 2.1 (and BB99, BB04) we write down a set of coupled conservation equations:    
\be
\label{angmom2}
{d\over dr}(G_1 - \dot M_1 L_1) = - G_{12}, 
\ee
\be
\label{angmom3}
{d\over dr}(G_2 + \dot M_2 L_2) = G_{12} ,
\ee
\be 
\label{energy2}
{d\over dr}(G_1 \Omega_1 - \dot M_1 B_1) =  - Q_{12} - \dot{\cal E}_1 ,
\ee
\be 
\label{energy3}
{d\over dr}(G_2 \Omega_2 + \dot M_2 B_2) =  Q_{12} - \dot{\cal E}_2,
\ee
 where $G_{12}$ and $Q_{12}$ represent the fluxes of angular momentum and energy per unit radius, respectively, transferred from zone 1 to zone 2, and $\dot{\cal E}_1$ and $\dot{\cal E}_2$ represent net radiative losses from each zone.  By mass conservation, the mass fluxes in zones 1 and 2 must be related according to $\dot M_1 (R) = \dot M_2 (R)+ \dot M_0$, where $\dot M_0$ is a constant representing the flux of gas that reaches the central object.  We can therefore drop the subscript on $\dot M_2$ and write $\dot M_1 = \dot M + \dot M_0$.

The angular momentum and energy fluxes transferred between zones cancel when the pairs of conservation equations are summed.  We can integrate the summed pairs of equations, to obtain
\be 
\label{FL}
\dot M (L_1 - L_2)= G_1 + G_2 + F_L - \dot M_0 L_1 ,
\ee
\be 
\label{FE}
\dot M (B_1 - B_2)= G_1 \Omega_1 + G_2 \Omega_2 - F_E + \int^R_{R_0} (\dot{\cal E}_1 + \dot{\cal E}_2) dR - \dot M_0 B_1 ,
\ee
where the constants of integration $- F_L$ and $F_E$ represent outward fluxes of angular momentum and energy, respectively, and we assume that the viscous stresses vanish at the innermost radius $R_0$.  Since we are interested in radii far outside the inner boundary, where $\dot M \gg \dot M_0$, we can neglect the terms containing $\dot M_0$ in both equations.  Since $\dot M L$ is an increasing function of $R$, we can neglect $F_L$ as well.   Since the viscous stress is assumed to be directed outward, eq.~(\ref{FL}) then implies that $L_1 > L_2$, i.e., that the inflow is spinning more rapidly than the outflow at each radius.  

In the one-zone ADIOS model, BB99 and BB04 neglected $F_E$ as well, which is justifiable for self-similar solutions with $n < 1$.  In the two-zone case, however, this assumption yields the curious result that $B_1 > B_2$, i.e., that the inflowing gas is {\it less} bound than the outflowing gas. Since the motivation for considering ADIOS models is the tendency of accreting gas to become unbound if it has no way to give up energy or angular momentum, this result seems suspect.  Indeed, we shall show below that it is also incompatible with the condition of dynamical equilibrium.  
 
\subsection{Self-consistency of two-zone models}
\label{ssec:cons}

To assess the self-consistency of the two-zone models, we introduce a generalized version of the dynamical equilibrium condition.  We need to include inertial forces in the outflow zone, where the speed ($v_2$) may be close to sonic.  We continue to assume that the inflows are driven by viscous stresses at (presumably) very subsonic speeds.   When mass is being transferred from the inflowing zone to the outflowing zone, the introduction of slowly moving gas emulates a drag force in the outflow region.  For a power-law model with $\dot M \propto R^n$, the resulting deceleration is given roughly by  
\be 
\label{inertial}
- {v_2 \over 2\pi R \Sigma_2}  {\partial \dot M\over \partial R} =   {-n v_2^2\over R}, 
\ee 
where we ignored the speed of the inflowing gas. The radial momentum equation for zone 2 then takes the form 
\begin{equation}
\label{mom2}
(1/2 - n)v_2^2 + {L_2^2 \over R^2}-{1\over R}+(5/2-n) a_2^2 =0,
\end{equation}
where we have defined an isothermal sound speed $a = (P/\rho)^{1/2}$.  The radial momentum equation for zone 1 takes the form 
\begin{equation}
\label{mom1}
{L_1^2 \over R^2}-{1\over R}+(5/2-n) a_1^2 =0,
\end{equation}
similar except for the neglect of dynamical terms. As in BB04, we ignore the angular correction terms associated with the finite scale height of each zone, effectively assuming that each zone is geometrically thin.  This will lead to errors of order $(H/R)^2$, where $H$ is the scale height. By comparing the one-dimensional models to two-dimensional (gyrentropic) models, BB04 showed that the 1D models provide an excellent approximation, at least for the inflow zones.  Since the outflow zones sandwich the inflow, we are also implicitly assuming that the $H_2$ is substantially larger than $H_1$. 

As in the one-zone models, we can eliminate one of the variables in each zone in favor of the Bernoulli function.  We modify the Bernoulli function for zone 2 to include the radial kinetic energy:
\be 
\label{Ber2}
B_2 = {v_2^2 \over 2} + {L_2^2\over 2 R^2} - {1\over R} + {\gamma\over \gamma -1} a_2^2,
\ee
with a similar expression (absent the dynamical term) for zone 1.  If we subtract eq.~(\ref{mom2}) from eq.~(\ref{mom1}) and eliminate $a_1^2 - a_2^2$, we obtain
\begin{eqnarray} 
\label{delB1}
B_1 - B_2 &=& -{1\over R^2}{\left({5\over 2} - n\right) - \gamma \left({1\over 2} -n \right) \over 2(\gamma -1)\left({5\over 2}-n \right)} \left( L_1^2 - L_2^2 \right) \nonumber \\
&& \qquad\qquad + {\left({5\over 2} - n\right)- \gamma \left({3\over 2} + n \right) \over 2(\gamma -1)\left({5\over 2}-n \right)} v_2^2 .
\end{eqnarray}
Since the coefficient multiplying $L_1^2 - L_2^2$ is always negative, eq.~(\ref{delB1}) shows that $B_1 - B_2$ and $L_1 - L_2$ must have {\it opposite} signs if dynamical terms are negligible.  This contradicts the conservation laws, equations (\ref{FL}) and (\ref{FE}), which demand that these quantities have the same sign in the zero-flux ($F_E = 0$) limit.  

In order to change the sign of $B_1 - B_2$, the term proportional to $v_2^2$ must be sufficiently positive and large. A necessary condition is 
\be 
\label{nlimit}
n < { 5-3\gamma \over 2 (\gamma + 1)},
\ee
which restricts $n$ to values between 0 (for $\gamma = 5/3$) and $1/2$ (for $\gamma = 1$).  To check whether consistency is possible even for such low values of $n$, we use the momentum equation to eliminate $v_2^2$ in favor of $a_1^2 - a_2^2$.  We obtain 
\begin{eqnarray} 
\label{delB2}
B_1 - B_2 &=& -{1\over R^2}{n + {1\over 2} \over 2 \left({1\over 2}-n \right)} \left( L_1^2 - L_2^2 \right) \nonumber \\
&& \qquad + {\left({5\over 2} - n\right)- \gamma \left({3\over 2} + n \right) \over (\gamma -1)\left({1\over 2}-n \right)} \left(a_1^2 - a_2^2 \right) .
\end{eqnarray} 
Since $n < 1/2$, the coefficient of $L_1^2 - L_2^2$ is still negative, implying that the other term must be positive.  But in the coefficient of $a_1^2 - a_2^2$ both the numerator and denominator are positive, implying that $a_2 < a_1$.  This would mean that the outflow must be ``colder" than the inflow, which seems unlikely given that it lies on top of the inflow, is in vertical pressure balance with it, and must have a higher specific entropy since it consists of adiabatic gas that has already passed through the inflow and has suffered dissipative heating in the process. 

Thus, it appears that the constraints of {\it dynamical} equilibrium in a two-zone ADIOS are incompatible with the constraints of energy and angular momentum conservation, if the net energy flux passing through the system is negligible.  This bolsters our original argument, based on the implausibility of $B_1 - B_2$ being positive on energetic grounds. 

How is this paradox resolved? The only reasonable conclusion seems to be that $F_E \neq 0$ and that a two-zone ADIOS model requires a substantial flux of energy passing through the system from small radii. If radiative losses are negligible, this flux is conserved.  But in the presence of radiative losses, this flux must be large enough to both change the sign of $B_1 - B_2$ at small radii and compensate for the integrated radiative losses further out.  In the limit of negligible radiative losses, this implies that $\dot M B$ is independent of radius and $n$ is not a free parameter, but must instead be $\approx 1$.

\section{Nonradiative Two-zone ADIOS}

We first consider the case in which radiative losses can be neglected, and therefore set $\dot{\cal E}_1 = \dot{\cal E}_2 = 0$.  Compared to the one-zone ADIOS model (BB04), the two-zone model would appear to be more restrictive since $n=1$ and $\beta =0$.  However, two-zone ADIOS models still form a three-parameter family because the energy flux $F_E$ is split into a portion that travels through the inflow and a portion that travels through the outflow. Defining 
\be 
\label{g12}
G_{12} = (1 + \eta) L_1 {d \dot M \over d R},
\ee
we can take $\eta$ to be a third parameter representing the ``lever arm" associated with angular momentum transfer, with a permissible range $0\leq \eta < 1/2$.  On physical grounds, we expect values of $\eta > 0$ to apply if organized magnetic fields mediate the transfer of angular momentum (BB04).

Denote the dimensionless energy flux by 
\be 
\label{mudef}
\mu \equiv {R F_E \over \dot M}
\ee
(where we recall that ${\rm G}M =1$) and let $\varepsilon$ be the fraction of the flux that propagates through the outflow. If we define the dimensionless variables
\be 
\label{dimensionless}
g\equiv {G\over \dot M R^{1/2}} ; \qquad \ell \equiv {L\over R^{1/2}} = R^{3/2} \Omega; \qquad b\equiv R B,
\ee
with subscripts 1 or 2 as appropriate, then the integrated conservation equations (\ref{angmom2})--(\ref{energy3}) become
\begin{eqnarray}
\label{dimlesseqns}
g_1 = {1\over 3} (1 - 2\eta) \ell_1 &;& \qquad g_2 = {2\over 3} (1+\eta) \ell_1 - \ell_2 \\
g_1 \ell_1 = b_1 + (1-\varepsilon) \mu  &;& \qquad g_2\ell_2 = -b_2 + \varepsilon \mu. 
\end{eqnarray}
We therefore have four conservation equations relating the 6 dimensionless variables $g_1, \ g_2, \ \ell_1, \ \ell_2, \ b_1$, and $b_2.$ Note that $\dot M$ is not an independent variable, since solutions are invariant under transformations that keep $G/\dot M$ and $F_E/ \dot M$ fixed. In addition, we have two dynamical equilibrium conditions, which we can cast in terms of the Bernoulli function by eliminating the sound speed $a$ and introducing a dimensionless radial speed $u_2 \equiv R^{1/2} v_2$: 
\begin{eqnarray}
\label{dimlesseqn1}
b_1 &=& {3 - \gamma \over 3 (\gamma -1)} \left[ 1 - {3 + \gamma \over 2 (3 - \gamma)} \ell_1^2 \right] \\
\label{dimlesseqn2}
b_2 &=& {3 - \gamma \over 3 (\gamma -1)} \left[ 1 - {3 + \gamma \over 2 (3 - \gamma)} \ell_2^2 \right] + {5\gamma - 3 \over 6 (\gamma - 1)} u_2^2. 
\end{eqnarray}

We thus have 6 equations for the 7 variables (including $u_2$), which formally define a line through the $(\mu, \varepsilon, \eta)$ parameter space. However, we can make the assumption that the viscous stress, in either zone, is capable of driving only very subsonic radial velocities. Therefore, if $g_2$ is non-negligible, then $u_2$ is negligible; conversely, a dynamically significant outflow speed implies that $g_2 \approx 0$, for our purposes.  We can therefore consider two distinct cases, for each of which we have 6 variables and 6 conditions.  We refer to these two cases as ``wind" solutions ($g_2 =0$) and ``breeze" solutions ($u_2 = 0$), respectively.  Given a complete set of parameters $(\mu, \ \varepsilon, \ \eta)$, the two-zone ADIOS model should be completely determined in each limit. 

\subsection{Wind solutions}

As the wind solution is slightly simpler algebraically, we start by analyzing it. Setting $g_2 = 0$ in the dimensionless equations, we immediately obtain
\be 
\label{b2wind}
b_2 = \varepsilon \mu \ ; \qquad \ell_2 = {2\over 3} (1+\eta) \ell_1.
\ee
Since $b_2 > 0$, the outflow is always unbound in the wind limit. 

The equations for the inflow region are self-contained, i.e., they do not depend on quantities from zone 2, implying that the same solution applies for both the wind and breeze cases. The Bernoulli function in zone 1 is given by eq.~(\ref{dimlesseqn1}), but an alternate expression can be derived by eliminating $g_1$ between the energy and angular momentum equations:
\be 
\label{b1wind}
b_1 = {1\over 3} (1 - 2\eta)\ell_1^2 -  (1-\varepsilon) \mu  .
\ee
Comparing the two expressions, we obtain a simple expression for $\ell_1^2$ in terms of the input parameters:
\be 
\label{ell1}
\ell_1^2 =  2 \ {3-\gamma + 3 (\gamma-1)(1-\varepsilon)\mu \over 3 + \gamma + 2(\gamma-1)(1-2\eta)}.
\ee
Given $\ell_1^2$, we can substitute from eq.~(\ref{b2wind}) into eq.~(\ref{dimlesseqn2}) and solve for $u_2^2$.

\begin{figure}
  \centering
  \includegraphics[bb=0 0 479 487,width=3.24in,height=3.29in,keepaspectratio]{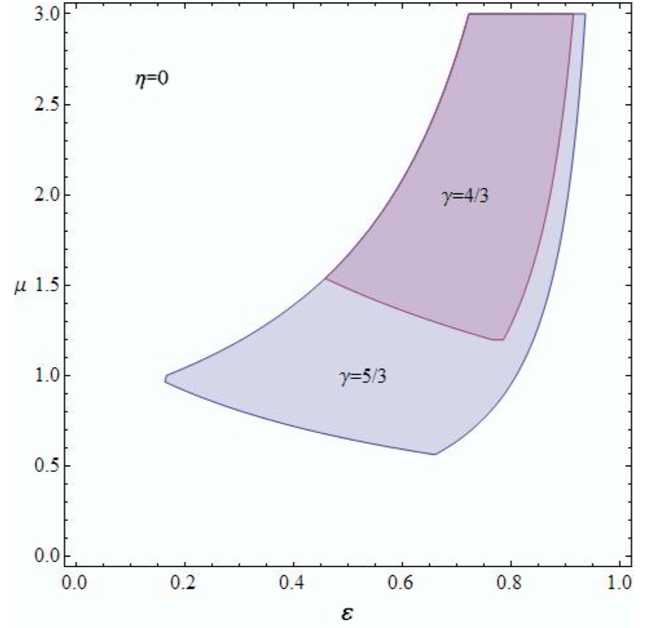}
  \caption{Parameter space of wind models in the $\varepsilon - \mu$ plane for $\gamma = 4/3$ and $\gamma=5/3$, with the angular momentum-transfer parameter $\eta$ set to zero. Here and elsewhere, $\mu$ is the outward energy flux normalized to ${\rm G}M\dot M/R $ and $\varepsilon$ is the fraction of that flux passing through the outflow zone.  For each shaded region, the left boundary corresponds to an inflow zone with Keplerian rotation (zero-pressure), while the right boundary corresponds to marginally bound inflow ($b_1 = 0$). The lower boundary of each region corresponds to the limit of vanishing outflow speed ($u_2 = 0$).  Below this boundary, the outflow must be driven by viscous stress and the system is in the ``breeze" limit.}
  \label{fig:wind5343}
\end{figure}

\begin{figure}
  \centering
  \includegraphics[bb=0 3 381 387,width=3.2in,height=3.22in,keepaspectratio]{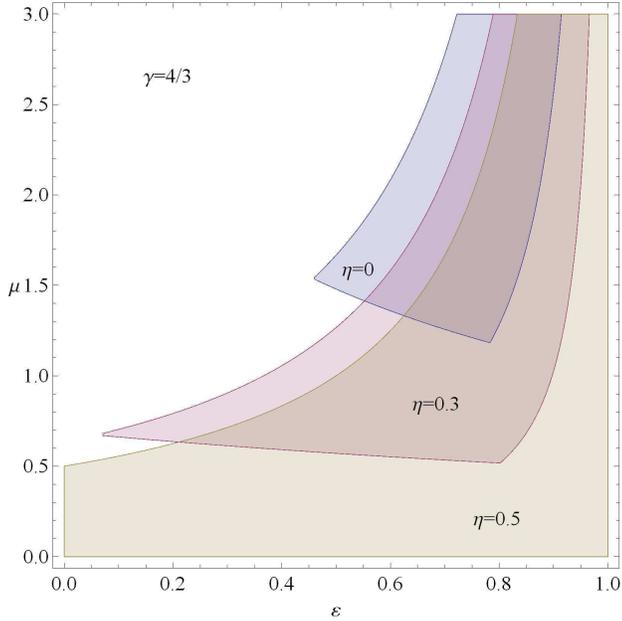}
  \caption{Parameter space for wind models with $\gamma=4/3$ and different values of the angular momentum-transfer parameter $\eta = 0, \ 0.3, \ 0.5 $.  The more efficiently angular momentum is transferred from inflowing to outflowing gas, the lower the central energy flux that is required to drive a wind.}
  \label{fig:windeta43}
\end{figure}

We are now able to map out the parameter space for wind solutions.  The three constraints on physical solutions are:
\begin{itemize}
  \item $\ell_1^2 < 1$, to ensure that the rotation is sub-Keplerian and the pressure (sound speed) is positive;
  \item $b_1 < 0$, to ensure that the inflowing gas is bound; and
  \item $u_2^2 > 0$, a self-consistency condition on the wind solutions.
\end{itemize}
These constraints are plotted in Figures \ref{fig:wind5343} and \ref{fig:windeta43}.

Figure \ref{fig:wind5343} shows the parameter space available to wind solutions in the $\varepsilon-\mu$ plane, for two values of $\gamma$.  The angular momentum-transfer parameter $\eta$ is set to zero, corresponding to the purely local transfer of angular momentum from inflowing to outflowing gas, i.e., there is no extra moment arm in the transfer due, for example, to an organized magnetic field (BB04).  The permitted regions are bounded on the left and right by the conditions $\ell_1^2 = 1$ and $b_1 = 0$, respectively.  This ordering is somewhat counterintuitive, because $\varepsilon$ represents the fraction of net energy flux that passes through the {\it outflow}. For a fixed total energy flux $\mu$, a larger value of $\varepsilon$ means less energy flowing through zone 1, yet the trend is for the binding energy of zone 1 to {\it decrease} with increasing $\varepsilon$.  Equally counterintuitive, we note that for fixed $\varepsilon$, an increase of total energy flux $\mu$ drives zone 1 from unbound flow toward a thin, cold state. The lower limit of each permitted region corresponds to zero outflow speed $u_2$, clearly a constraint on the validity of the wind model.  The fact that the allowed parameter space for $\gamma=4/3$ is more restictive than for $\gamma = 5/3$ indicates that it requires more energy flowing through the system (and a larger fraction of it flowing through the outflow zone) to drive an outflow in $\gamma = 4/3$ gas.  This is not surprising, given that $\gamma = 4/3$ is the softer equation of state.

Figure \ref{fig:windeta43} shows the effect of varying $\eta$, with $\gamma = 4/3$.  As the efficiency of angular momentum transfer increases, less energy flux (and a smaller fraction flowing through the outflow) is required to drive a wind.  in the limiting case $\eta = 1/2$, winds can be driven without any net energy flux through the system.

\subsection{Breeze solutions}

To study the breeze solutions, we express $b_2$ in terms of $\ell_2$ by setting $u_2^2 =0$ in eq.~(\ref{dimlesseqn2}). Eliminating $g_2$ between the angular momentum and energy equations for zone 2, we obtain a second equation for $b_2$, in terms of $\ell_2$ and $\ell_1$.  Eliminating $b_2$ between these equations, we obtain a quadratic equation for $\ell_2$:
\be 
\label{ell2quad}
{7\gamma-3 \over 6(\gamma-1)} \ell_2^2 - {2\over 3} (1+\eta) \ell_1\ell_2 + \varepsilon \mu - {3 - \gamma\over 3(\gamma -1)} = 0,
\ee
the solution of which is
\begin{eqnarray}
\label{ell2quad2}
\ell_2 &&=     (1+\eta){2 (\gamma-1)\over 7\gamma-3} \Biggl[ \ell_1  \hfill \nonumber \\
&& \left. \pm \sqrt{\ell_1^2 + {3\over 2}{7\gamma - 3 \over (\gamma-1) (1+\eta)^2}\left( {3 - \gamma\over 3(\gamma -1)} - \varepsilon \mu \right) } \right].
\end{eqnarray}
Since $\ell_1$ is still given by eq.~(\ref{ell1}), we can solve for all the properties of the breeze solutions in terms of the input parameters. 

\begin{figure}
  \centering
  \includegraphics[bb=0 3 381 387,width=3.2in,height=2.89in,keepaspectratio]{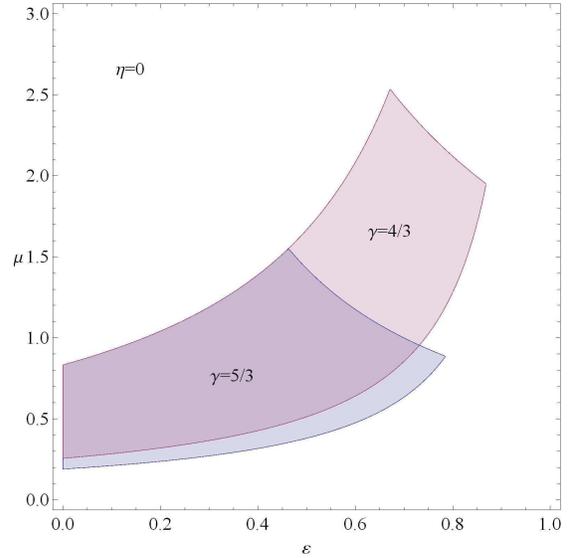}
  \caption{Parameter space of breeze models in the $\varepsilon - \mu$ plane for $\gamma = 4/3$ and $\gamma=5/3$, with the angular momentum-transfer parameter $\eta$ set to zero. For each shaded region, the left boundary corresponds to an inflow zone with Keplerian rotation (zero-pressure), while the right boundary corresponds to marginally bound inflow ($b_1 = 0$) --- as in the wind solutions. The upper boundary of each region corresponds to the limit of vanishing discriminant in eq.~(\ref{ell2quad2}).  Above this boundary, no breeze solution exists.}
  \label{fig:breeze5343}
\end{figure}

\begin{figure}
  \centering
  \includegraphics[bb=0 3 381 387,width=3.2in,height=2.71in,keepaspectratio]{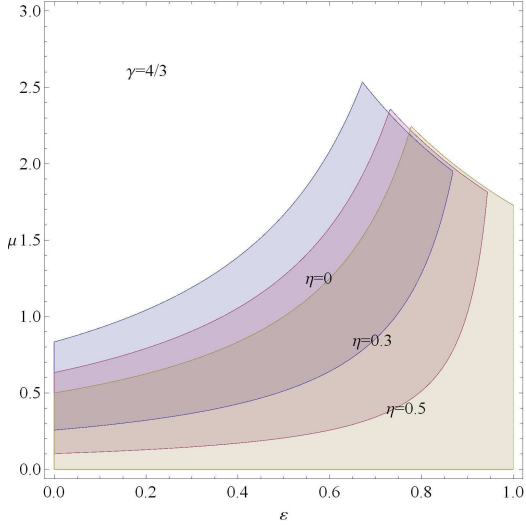}
  \caption{Parameter space for breeze models with $\gamma=4/3$ and different values of the angular momentum-transfer parameter $\eta = 0, \ 0.3, \ 0.5 $.  The overlapping bands show the same trend as for the wind models, i.e., that a higher angular momentum transfer efficiency reduces the energy flux needed to drive a breeze. Note that the upper-right threshold for the parameter space is very insensitive to the value of $\eta$.}
  \label{fig:breezeeta43}
\end{figure}

Breeze solutions are subject to four constraints:
\begin{itemize}
  \item $\ell_1^2 < 1$ as before, to ensure that the rotation of the inflowing gas is sub-Keplerian and the pressure (sound speed) is positive;
  \item $b_1 < 0$, also as before, to ensure that the inflowing gas is bound;
  \item $g_2 > 0$, the assumption that the stress points outward in the outflow, also ensuring that $\ell_2 < {2\over 3} (1+\eta) \ell_1 < 1$; and
  \item $\ell_2$ has real solutions according to eq.~(\ref{ell2quad2}), which serves as a self-consistency condition for the breeze solutions. 
\end{itemize}
From eq.~(\ref{ell2quad2}), one can easily see that the condition $g_2 > 0$ is well-satisfied when the discriminant in the expression for $\ell_2$ vanishes.  One can also show that this condition continues to be satisfied at all values of the parameters for which a breeze solution exists.  Therefore, the third condition is superfluous and the parameter space of viable breeze solution is determined by the conditions on $\ell_1$, $b_1$, and the existence of a real solution for $\ell_2$.  These constraints are plotted in Figures \ref{fig:breeze5343} and \ref{fig:breezeeta43}. All breeze solutions have unbound outflow, $b_2 > 0$.

As before, each solution space occupies a strip between the Keplerian limit at the left and the limit of marginally bound inflow at the right.  The upper boundary marks the locus on which the discriminant of the quadratic equation (\ref{ell2quad}) vanishes.  No physical breeze solution exist for parameters above this boundary.  By comparing Fig.~\ref{fig:breeze5343} with Fig.~\ref{fig:wind5343}, one can also note that the parameter space for breeze solutions overlaps with the parameter space for winds; thus, in principle one could have either a breeze or a wind model for the same set of parameters $\mu, \ \varepsilon, \eta$.  Figure \ref{fig:breezeeta43} shows the effect of varying $\eta$, with $\gamma = 4/3$.  As in the wind case, a higher efficiency of angular momentum transfer decreases the energy flux required to drive a wind.   

\subsection{Combined parameter space}

\begin{figure}
  \centering
  \includegraphics[bb=0 3 381 387,width=3.2in,height=3.2in,keepaspectratio]{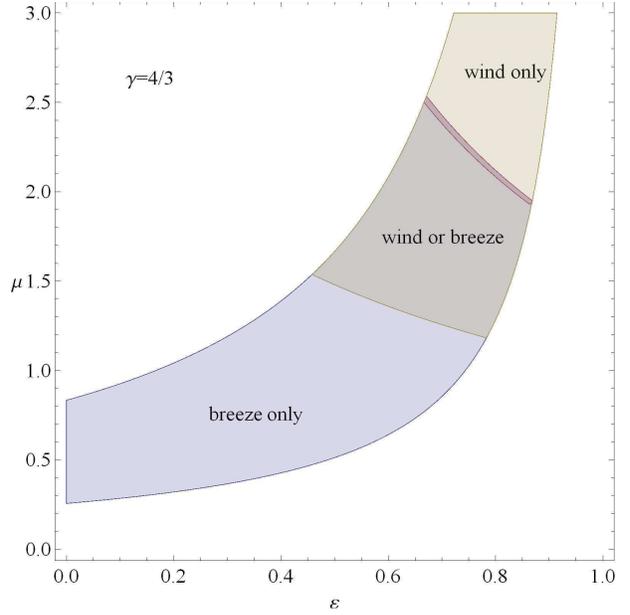}
  \caption{Combined $\mu-\varepsilon$ parameter space of solutions for $\gamma=4/3$, with $\eta=0$. Both wind and breeze solutions may exist in the dark shaded area.  Slowly rotating breeze solutions (i.e., with the minus sign in eq.~[\ref{ell2quad2}]) may exist only in the narrow strip between the wind-only area and the ``wind or breeze" region; elsewhere only rapidly rotating breeze solutions are allowed.}
  \label{fig:windbreeze43}
\end{figure}

\begin{figure}
  \centering
  \includegraphics[bb=0 3 381 387,width=3.2in,height=3.2in,keepaspectratio]{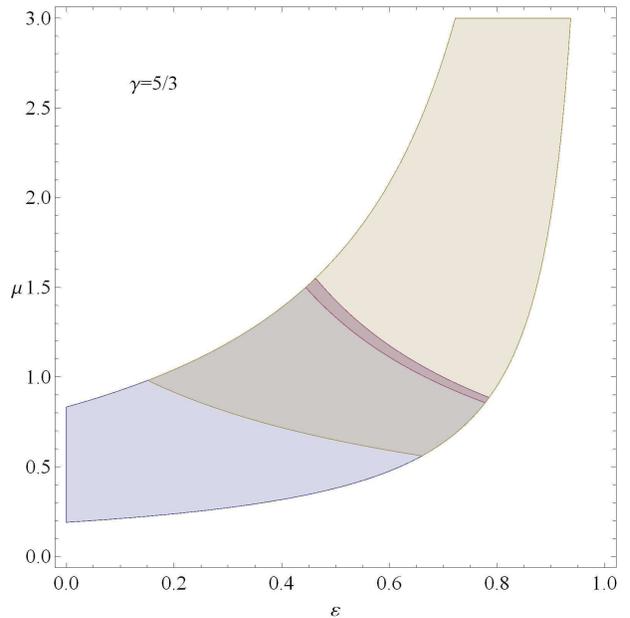}
  \caption{Same as Figure \ref{fig:windbreeze43}, but for $\gamma=5/3$.}
  \label{fig:windbreeze53}
\end{figure}

We are now in a position to combine our results for the breeze and wind models, to show the different possible types of solution and how they relate to the input parameters.  In Figures \ref{fig:windbreeze43} and \ref{fig:windbreeze53} we show the combined solution space for $\gamma=4/3$ and $\gamma=5/3$, respectively.  The first thing to notice is that wind and breeze regions overlap substantially.  Moreover, it is possible that both rapidly rotating (plus sign in eq.~[\ref{ell2quad2}]) and slowly rotating breeze solutions may be possible, although the slowly rotating solutions are permitted only in a very narrow band of parameter space. (This constraint is based on the assumption that the outflow rotates in the same sense as the inflow.)

These results provide a complete picture of the two-zone ADIOS models when losses are neglected.  We will now consider models that include self-similar radiative losses.  

\section{Two-zone ADIOS with Radiative Losses}

Radiative losses are included on the right-hand side of equations (\ref{energy2}) and (\ref{energy3}) by specifying the form of $\dot{\cal E}_1$ and $\dot{\cal E}_2$.  To guarantee self-similar scaling for these losses, we follow Narayan \& Yi (1994) and define a parameter $f$, with $0 < f \leq 1$, such that $- f G d\Omega/ dR$ equals the net heating rate per unit radius.  Since the total viscous heating rate per unit radius is  $- G d\Omega/ dR$, we have $\dot{\cal E}_1 = - (1-f_1) G_1 d\Omega_1/dR$, with a similar expression for zone 2.  Note that $f=1$ corresponds to the nonradiative case we have been considering, while $f\rightarrow 0$ implies that all locally dissipated energy is radiated away.  We also specify $Q_{12} = (1-\beta) B_1 (d\dot M/dR)$, as in BB04. 

We define the dimensionless constants $b$ and $\ell$ as in eq.~(\ref{dimensionless}), but now define $G = \tilde g R^{3/2}$ and $\dot M = \dot m R$ such that $\tilde g$ and $\dot m$ are functions of $R$. The dimensionless form of the cooling rate then has the form $\dot{\cal E}_1 R = 3 (1-f_1)\tilde g_1 \ell_1/2$ in zone 1, with a similar form for zone 2.

Under these assumptions, the conservation equations admit power-law solutions of the form
\be 
\label{radsolns}
\left\{\dot m, \tilde g_1, \tilde g_2 \right\} = \left\{ A,  C_1, C_2 \right\} R^{-q},
\ee
where $A$,  $C_1$, $C_2$, and $0 < q < 1$ are constants. We may choose the overall normalization arbitrarily, so we set $A=1$. 

Substituting eq.~(\ref{radsolns}) into the conservation equations, we can obtain solutions for arbitrary choices of $\eta$, $f_1$ and $f_2$, but for the purpose of this section we can simplify the algebra considerably by considering the case $\eta = 0$ with $f_1 = f_2 = f$. We then have
\be 
\label{C1}
C_1 = {\ell_1 \over 3 - 2 q}
\ee 
and
\be 
\label{C2}
C_2 = \ell_1 - \ell_2 - C_1
\ee 
Defining $y \equiv 3 - 2q$, we obtain
\be 
\label{yeq}
(y - 3f)(\ell_1 - \ell_2)(\ell_1 + y \ell_2) + y (y-3)(b_2 - b_1) = 0,
\ee     
which appears to be a quadratic equation for $y$ (or $q$) but is actually cubic because $b_1$ and $b_2$ depend linearly on $q$.  To see this, we modify the momentum equations (\ref{mom2}) and (\ref{mom1}), replacing $n$ by $1-q$.  We then have  
\begin{eqnarray}
\label{b1q}
\lefteqn{b_1  =   {1 \over (\gamma -1)(3+ 2q)}  \left[ 3 - \gamma - 2q(\gamma-1) \right.  \nonumber } \\
    & & \qquad\qquad\qquad\qquad \left. - {3 + \gamma -2q (\gamma-1)\over 2} \ell_1^2 \right] \\
\label{b2q}
\lefteqn{b_2  =   {1 \over (\gamma -1)(3+ 2q)} \left[ 3 - \gamma - 2q(\gamma-1) \right.  \nonumber } \\
  & & \left. - {3 + \gamma -2q (\gamma-1)\over 2} \ell_2^2 \right]+{5\gamma - 3 -2q(\gamma -1) \over 2 (\gamma - 1)(3+2q)} u_2^2 . 
\end{eqnarray}
Given $\ell_1$ and $q$, we can calculate $\beta$ from the equation
\be 
\label{betaq}
1- \beta = {1\over 1-q} \left[ q - {\ell_1^2 \over  b_1} {3(1-f) -2q \over 2 (3-2q)} \right],
\ee
which derives from eq.~(\ref{energy2}).

These formal solutions to the equations suffer from the self-consistency problem described in section 2.3 unless the system carries a central energy flux $F_E >0$ to partially compensate for radiative losses.  From eq.~(\ref{FE}), however, we see that the presence of the energy flux would spoil the assumed power-law dependence unless 
\be
\label{FErad}
F_E = \int^\infty_{R_0} (\dot{\cal E}_1 + \dot{\cal E}_2) dR.
\ee
In such a {\it critically cooled} solution, the system ``forgets" about the energy flux injected at the center and instead carries an $R$-dependent effective energy flux 
\be
\label{FErad2}
 \int^\infty_R (\dot{\cal E}_1 + \dot{\cal E}_2) dR,
\ee
which allows self-consistency to be maintained.  We further note that if $F_E$ were smaller than the energy integral in eq.~(\ref{FErad}), then no solution would be possible --- cooling would exceed the total energy supply.  Presumably, any radiative system will adjust so that the accretion rate yields a central energy flux at least large enough to balance cooling.  Conversely, if $F_E$ exceeded the energy integral, then radiative losses would have a minor effect and the nonradiative ($n=1$) behavior would be preserved.

Below we consider critically cooled models in the breeze and wind limits.

\subsection{Breeze solutions}  

To extract the breeze solutions, we set $u_2^2 =0$ in eq.~(\ref{b2q}) but make no assumption about $\tilde g_2$.  Substituting for $b_2 - b_1$ in eq.~(\ref{yeq}) and canceling a factor of $\ell_1 - \ell_2$, we obtain the cubic equation
\be 
\label{yeq2}
(y - 3f)(\ell_1 + y \ell_2) + y (y-3){3 + \gamma - 2q(\gamma-1)\over 2 (\gamma -1)(3+ 2q) }(\ell_1 + \ell_2) = 0,
\ee 
where we recall that $y = 3 - 2q$. Breeze solution must satisfy the constraints:
\begin{itemize}
  \item $-1/2 < b_1 <0$;
  \item $\ell_2 >0$; and 
  \item $\tilde g_2 > 0$.
\end{itemize}
The last condition implies $\ell_2 < \ell_1 (1 - y^{-1})$.  Since $y-3 = -2q <0$, the second term on the left-hand side of eq.~(\ref{yeq2}) is negative and we must have $y -3f >0$, which translates to $q < 3(1-f)/2$.
If $b_1 < 0$ as well, then eq.~(\ref{betaq}) implies that 
\be 
\label{betaqmin}
1 - \beta > {q \over 1-q}.
\ee

\begin{figure}
  \centering
  \includegraphics[bb=1 2 431 412,width=3.32in,height=3.17in,keepaspectratio]{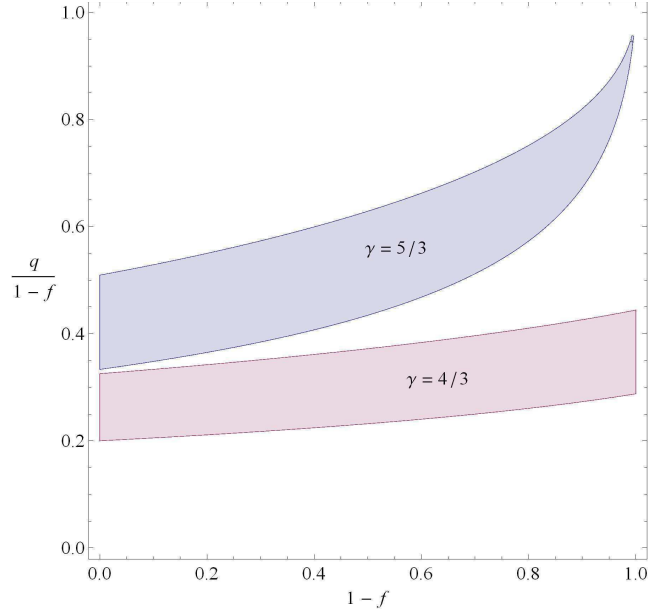}
  \caption{Allowed values of the exponent $q$ ($\dot M \propto R^{1-q}$), normalized to the radiative fraction $1-f$, as a function of $1-f$ for $\gamma = 4/3$ and $\gamma = 5/3$,.  For each case, the lower bound corresponds to vanishing angular momentum in the outflow, $\ell_2 = 0$, while the upper bound corresponds to vanishing viscous stress, $G_2 = 0$.  Note that $q$ vanishes $\propto 1-f$ in the limit of negligible radiative losses, thus recovering the case $n=1$ ($\dot M \propto R$).}
  \label{fig:q}
\end{figure}

The allowed range of $q(f)$ is determined by the second and third constraints. The actual value of $\ell_1$ (or $b_1$) is irrelevant, since from eq.~(\ref{yeq2}) we see that $q$ is a function of $\ell_2/\ell_1$, not of either angular momentum separately.  The second and third constraints imply $0 < \ell_2/\ell_1 < 1- y^{-1}$; we map this region in Fig.~\ref{fig:q}. 

One can derive these constraints analytically in the limit of negligible radiative losses, $f\rightarrow 1$:
\be 
\label{qzero}
{3(\gamma - 1)\over 3\gamma+1} < {q\over 1-f} < {27 (\gamma -1)\over 23\gamma -3}.
\ee  
As the radiative fraction $1-f$ goes to zero, $q$ does as well and the solution approaches the case $n=1$.

\subsection{Wind solutions}

For the wind case, we take $\tilde g_2 \approx 0$, implying $C_2 =0$. Combining equations (\ref{C1}) and (\ref{C2}), we obtain
\be 
\label{ell2q}
\ell_2 = {2\ell_1 (1-q) \over 3-2 q } ,
\ee 
which generalizes eq.~(\ref{b2wind}) for $\ell_2$ in the case $\eta = 0$.  Equation (\ref{yeq}) can then be written in the form
\be 
\label{ell2q2}
b_2 - b_1 =  {3(1-f) - 2q \over 2q (3-2q)} \ell_1^2 = {b_1\over q}\left[(\beta -1)(1-q)-q  \right], 
\ee 
where the second version of the equation comes from eq.~(\ref{energy3}) with $\tilde g_2 = 0$.  These equations are identical to eq.~(\ref{betaq}).  

Wind solutions must satisfy the constraints:
\begin{itemize}
  \item $-1/2 < b_1 <0$; and
  \item $u_2^2 > 0$.
\end{itemize}
The latter condition places an upper limit on $q(f)$ that is identical to the upper boundary for the breeze zone in Fig.~\ref{fig:q}, i.e., corresponding to $G_2 = 0$. The allowed zone for wind solutions extends down to $q=0$ for all $f$.  The parameter space for breezes therefore completely overlaps with that for winds, in the $q-(1-f)$ plane.

\section{Dissipation}

Our two-zone models of ADIOS flows have been based on the conservation laws of angular momentum and energy, supplemented by dynamical equilibrium conditions in both the inflowing and outflowing zones.  In addition, these zones must obey the laws of thermodynamics, which describe how energy is distributed among various forms.  Consideration of the thermodynamics of dissipative heating can impose additional constraints on the possible solutions.  Since we have already found that models are completely determined by the choice of three input parameters ($\mu, \ \varepsilon, \ \eta$) --- plus, under certain conditions, an assumption about whether the outflow is a wind or a breeze with either fast or slow rotation --- it is clear that any further restrictions may lead to constraints on the possible values of these parameters.  For example, it may not be possible for the constant energy flux through the flow, or its partition between inflowing and outflowing zones, to take on any value --- the allowed values may instead be dictated by thermodynamics.

We are basically concerned with estimating the rate of entropy generation within each zone of the flow.  Under self-similar conditions with $n=1$, the pressure scales with radius as $P\propto R^{-3/2}$ and the density scales as $\rho \propto R^{-1/2}$ implying that the specific entropy function scales as $P/\rho^\gamma \propto R^{-(3-\gamma)/2}$.  The entropy thus increases with decreasing radius, as expected for inflowing gas that undergoes viscous dissipation (but does not radiate) as it drifts inward.  However, the self-similar nature of our two-zone model implies that the same radial entropy scaling must also apply to the outflowing gas --- in other words, the specific entropy of the outflowing gas must decrease as the gas flows outward.  This entropy decrease in the outflowing zone indicates that the entrainment of low-entropy gas from zone 1 must outweigh the entropy increase due to local dissipation. 

These arguments serve as a warning of the care required in attempting to estimate dissipative effects.  Setting $s \equiv \ln(P/\rho^{\gamma})$, we write the vertically integrated entropy equation for zone 1 in the approximate form
\be 
\label{entropy}
-{a_1^2 \over \gamma -1} \dot M {\partial s_1 \over \partial R} + {2\pi R a_1^2 \over \gamma -1} \int_1 \rho v_z {\partial s \over \partial z} dz = -G_1 {d\Omega_1 \over dR} - {d Q\over dR}, 
\ee
where the integral is over zone 1 and we have glossed over such subtleties as the difference between the vertical average of the sound speed used here (recall that $a^2 = P/\rho$) and those used in the dynamical equilibrium conditions.  A similar equation applies in zone 2, except that the first term on the left-hand side has the opposite sign. For now we ignore the $dQ/dR$ term, which represents energy lost per unit radius due to conduction or radiation.  We set $2\pi R \rho v_z \approx d\dot M/dR$, and assume that the second term on the left-hand side can be approximated by 
\be 
\label{entropy2}
 {a_1^2\over \gamma -1}   {d\dot M \over dR} \Delta s, 
\ee
where $\Delta s$ is the difference in specific entropy between the top and bottom of the zone under consideration. Presumably $\Delta s > 0$ in both zones 1 and 2. Writing ${\partial s / \partial R} = -(3 - \gamma )/ 2R$ in both zones, redefining $a$ so it is dimensionless ($a^2 \rightarrow Ra^2$), and using the dimensionless variables defined earlier, we obtain
\be 
\label{entropy3}
\Delta s_1 =  { 3 (\gamma -1) \over 2}{g_1 \ell_1\over a_1^2} - {3-\gamma \over 2} 
\ee
\be 
\label{entropy4}
\Delta s_2 =  { 3 (\gamma -1) \over 2}{g_2 \ell_2\over a_2^2} + {3-\gamma \over 2}. 
\ee
Thus, we see that in order to preserve the self-similar scaling, there must be a large entropy jump across zone 2, but there need not be one across zone 1.  For the sake of making a crude estimate, we will assume that the entropy jump across zone 1 is negligible, and that all the entropy change due to mass transfer between zones occurs in zone 2.  In this limit, we have
\be 
\label{entropy5}
{g_1 \ell_1\over a_1^2} = {(1-2\eta)\ell_1^2 \over 2(1 - \ell_1^2)} = {3-\gamma \over 3 (\gamma -1) }, 
\ee
where we have used the equations of energy conservation and dynamical equilibrium in zone 1 to express the left-hand side in terms of $\ell_1$. The entropy jump across zone 2, which must be approximately equal to $s_2 - s_1$ if we assume that $\Delta s_1 \approx 0$, is likely to be even larger than the estimate given in eq.~(\ref{entropy4}), which only includes the $R\phi-$component of the stress.  In addition to the possibility of a significant $z\phi$ stress due to the vertical shear in angular velocity, there may be large $Rz$ stress.  This will certainly be a factor for the wind solutions, where we have already had to take into account the drag on the outflow due to entrainment of gas with little radial velocity (cf.~eq.~[\ref{mom2}]).   

Solving eq.~(\ref{entropy5}) for $\ell_1^2$, we obtain
\be 
\label{entropy6}
\ell_1^2 = {2 (3-\gamma)\over 2 (3-\gamma) + 3 (\gamma -1)(1-2\eta) } .
\ee
In the limit $\eta\rightarrow 1/2$, angular momentum loss carries off all the accretion energy and we are left with a thin, Keplerian disk.  For all other cases, viscous heating is important and $\ell_1^2 < 1$, as expected.  Using eq.~(\ref{dimlesseqn1}) to express the inflow Bernoulli function in terms of $\ell_1^2$, we obtain 
\be 
\label{entropy7}
b_1 = - {2 (3-\gamma)\eta \over 2 (3-\gamma) + 3 (\gamma -1)(1-2\eta) } .
\ee
For the case $\eta =0$, $b_1 =0$ and the disk is marginally bound.  In this limit, the locus of permitted values of $\mu$ and $\varepsilon$ would track the right-hand boundary of the solution space in figures \ref{fig:windbreeze43} and \ref{fig:windbreeze53}.  For $0 < \eta < 1/2$ the disk is bound and the locus lies inside the parameter space.  For $\eta \rightarrow 1/2$, the locus hugs the left-hand boundary of the solution space and the Bernoulli function has the Keplerian value $b_1 = - 1/2$.
 
It proves interesting to generalize our dissipative model to the case where $\Delta s_1$ is positive, e.g., due to mixing with the hotter gas in the outflow zone.  In that case, eq.~(\ref{entropy5}) is replaced by the inequality
\be 
\label{entropy8}
{(1-2\eta)\ell_1^2 \over 2(1 - \ell_1^2)} > {3-\gamma \over 3 (\gamma -1) }. 
\ee
Since the left-hand side is a monotonically increasing function of $\ell_1^2$, this means that additional entropy generation leads to a colder, more rapidly rotating inflow, and to a more negative binding energy.  This surprising result could explain why simulations of radiatively inefficient flows tend to exhibit inflow zones with clearly negative values of the Bernoulli function, rather than values hovering close to zero (e.g., Stone et al.~1999). 

The effect of radiative losses is to replace the dimensionless stress $g$ in equation (\ref{entropy3}) or (\ref{entropy4}) by $fg$, according to the self-similar model of radiative losses adopted in section 4.  For the inflow zone, we obtain 
\be 
\label{heating1}
\ell_1^2 = {2 (3-\gamma)\over 2 (3-\gamma) + 3 f(\gamma -1)(1-2\eta) } ,
\ee
confirming that increasing the losses (decreasing $f$) makes the inflow colder and the rotation faster.  The Bernoulli function is  
\be 
\label{entropy7}
b_1 = - { (3-\gamma)\left[1 - f (1-2\eta)\right] \over 2 (3-\gamma) + 3 f(\gamma -1)(1-2\eta) } .
\ee
Thus, as expected, an increase in radiative losses makes the inflow more bound.  On the $\mu-\varepsilon$ parameter space, this moves the possible solutions to the left of the zero-energy curve.  If $\mu$ is fixed, this means that $\varepsilon$ is smaller, corresponding to a smaller fraction of the energy flux passing through the outflow.

\section{Hypercritical accretion}

To apply these ideas in a well-defined context, we consider the case of hypercritical accretion, in which matter is supplied at a rate that far exceeds the Eddington limit.  Such a flow is radiatively inefficient because of the large optical depth; indeed, the radiation becomes effectively ``trapped" in the flow (advected faster than it can escape) at roughly the radius within which the liberated power approaches the Eddington limit (Begelman 1979).

Radiative losses are determined using the vertical ($z$) component of the momentum equation.  Since radiation from zone 1 has to pass through zone 2, we set $\dot {\cal E}_1 = 0$ in eq.~(\ref{energy2}) and absorb the radiative losses from zone 1 into the overall energy loss term $Q_{12}$.  Net radiative losses are thus given by $\dot {\cal E}_2$.  In considering the vertical dynamical equilibrium of zone 2 we must take account of the outflow speed, which can be appreciable.  Modeling the vertical velocity profile by $v_{2z} = v_2 z/R \propto z/R^{-3/2}$, we find that the dimensionless vertical acceleration is $- u_2^2(z/R)/2$.  There is also a vertical drag force, associated with lifting material out of zone 1 and mixing it with the outflowing material in zone 2 (cf.~eq.~[\ref{inertial}]), that gives a deceleration $-u_2^2 (z/R)$.  Inserting these terms into the momentum equation, we obtain   
\be 
\label{Edot1}
\dot {\cal E}_2 = {4\pi {\rm G}M c\over \kappa R} \left( 1 + {u_2^2\over 2}\right){H_2 \over R} 
\ee
where $H_2$ is the height of the outflow layer, $\kappa$ is the opacity, and we include the ${\rm G}M$ factor for clarity (elsewhere we have taken ${\rm G}M = 1$). We have doubled the losses from one layer to take account of the fact that two radiating outflow layers sandwich the inflow zone.  To estimate $H_2$, we integrate the vertical momentum equation assuming a uniform density.  This gives $H_2/R = \sqrt{2} a_2/ (1 + u_2^2/2)^{1/2}$, where $a_2$ is the dimensionless isothermal sound speed on the equator.  We therefore have
\be 
\label{Edot2}
\dot {\cal E}_2 = {4\pi {\rm G}M c\over \kappa R} \sqrt{2} a_2 \left( 1 + {u_2^2\over 2}\right)^{1/2} . 
\ee
 
We define a dimensionless accretion rate $\dot m$ by $\dot M = 4\pi c R \dot m / \kappa$ and assume that matter is supplied at some large radius $R_{\rm out}$ at a rate $\dot M_{\rm out} = 4\pi c R_{\rm out} \dot m_{\rm out} / \kappa$.  If $R_{\rm out}$ is taken to be the radius at which radiation is marginally trapped by the inflow, then $\dot m_{\rm out} \sim O(1)$.  If the inner radius $R_{\rm in} \ll R_{\rm out}$, then it is possible to develop a series solution in powers of a small parameter $\delta$ at radii $R \ll R_{\rm out}$.  Importantly, the dimensionless coefficients we have used to characterize the flows remain constant to first order in $\delta$ --- they vary with radius only at $O(\delta^2)$.  In contrast, the mass flux and viscous stress develop a logarithmic radial behavior at first order in $\delta$, which truncates the ADIOS at $R_{\rm out}$. 

To see how this behavior arises, we develop solutions to first order for zone 1, anticipating the radial dependence of $\dot m$ (which is governed by the form of $\dot {\cal E}_2$) by writing
\begin{eqnarray}
\label{mdota}
\dot m &=& \dot m_a \left[ 1 -  \delta \ln\left( {R\over R_{\rm in}} \right) \right] \\
 G_1 &=& \dot m_a R^{3/2}\left[g_{1a} - g'_{1a}  \delta  \ln\left( {R\over R_{\rm in}} \right) \right] ,
\end{eqnarray} 
where quantities denoted by a subscript ``$a$" are regarded as zeroth order and quantities with a subscript 
``$b$" are first order. We correspondingly define $\ell_1 = \ell_{1a} + \ell_{1b}$ and $b_1 = b_{1a} + b_{1b}$.  Substituting these forms into the conservation equations, we recover equations (\ref{dimlesseqns}) and (23) as zeroth-order relations.  To first order in the angular momentum equation, the terms proportional to $\ln (R/R_{\rm out})$ give $g_{1a} = g'_{1a} = g_1$, implying that $G_1(R)= g_1 R^{3/2} \dot m(R)$.  The remaining first-order terms give the correction to the specific angular momentum,
\be
\label{ell1b}
\ell_{1b} = -{2\over 3}\delta{1+\eta \over 1- 2\eta}  \ell_{1a}.
\ee
Identifying $g_1 \ell_{1a} - b_{1a} = (1-\varepsilon) \mu$ as in the nonradiative case, we obtain
\be
\label{hyerbeta}
(1 - \beta ) b_{1a}  = \delta (1-\varepsilon) \mu .
\ee
In the presence of radiative losses, the energy transfer rate $Q_{12} \propto 1-\beta$ is no longer zero, but behaves as a first-order quantity.  

To proceed further, we need to use the perturbed version of the radial momentum equation.  As before, we have $p = \rho a^2$, so that $ p'/\rho = a^2(\rho'/\rho + 2 a' / a)$ with $a' / a = -1/2R$ to first order.  Since $\dot m \propto \rho v R^2$ with $v\propto a$ to first order, the density must contain the same first-order corrections as $\dot m$, and we have $R\rho'/\rho = -1/2 - \delta $.  The equilibrium condition is then   
\begin{equation}
\label{hypermom1}
\ell_1^2 - 1+\left({3\over 2} + \delta \right) a_1^2 =0,
\end{equation}
which we solve to first order after expanding $\ell_1^2$ and taking $a_1^2 \approx a_{1a}^2 + 2 a_{1a}a_{1b}$. Specializing to $\gamma = 4/3$ in the Bernoulli function, we obtain
\begin{eqnarray}
\label{hyperab}
\lefteqn{a_{1a}^2 + 2 a_{1a}a_{1b}  =   {2\over 3} (1 - \ell_{1a}^2) - {4\over 9} \delta \left(1 - {3\over 1-2\eta}\ell_{1a}^2 \right) } \\
\lefteqn{b_{1a} + b_{1b}  =   {5\over 3}\left(1 - {13 \over 10}\ell_{1a}^2 \right) - {16\over 9} \delta  \left(1 - {3\over 8}{7-\eta\over 1-2\eta}\ell_{1a}^2 \right)}.
\end{eqnarray}

A similar procedure can be carried out for zone 2, but here we quote only the important result obtained by comparing terms in the integrated, summed energy equation (\ref{FE}).  Noting that $G_2(R) \propto R^{3/2} \dot m(R)$ and that $F_E = 4\pi {\rm G}M c\dot m_a \mu /\kappa = \dot m_a\mu L_E$, where $L_E$ is the Eddington luminosity, we find that 
\be
\label{mudef}
\delta  =  {\sqrt{2} a_2 \over \dot m_a \mu} \left( 1 + {u_2^2\over 2}\right)^{1/2}.
\ee
The correction term in eq.~(\ref{mdota}) is negative, implying that the ADIOS is quenched at large radii by radiative losses.  In particular, if we extrapolate the logarithmic correction out to radii at which the correction is of order unity (recognizing that the approximation is a very crude one in this limit), we find that $\dot M$ has a maximum at a radius which we identify as $R_{\rm out}$, implying
\be
\label{Rout}
\delta^{-1} = 1 + \ln\left( {R_{\rm out} \over R_{\rm in}}\right) = {\dot m_a \over \dot m_{\rm out}}.
\ee
For exponentially large values of $R_{\rm out}/R_{\rm in}$, as we expect in hyperaccreting systems such as SS433 (where this ratio may be $\sim 10^3$), $\delta$ may fall in the range $\sim 0.1-0.2$, justifying our approximation scheme. 

\begin{figure}
  \centering
  \includegraphics[bb=0 0 440 426,width=3.24in,height=3.13in,keepaspectratio]{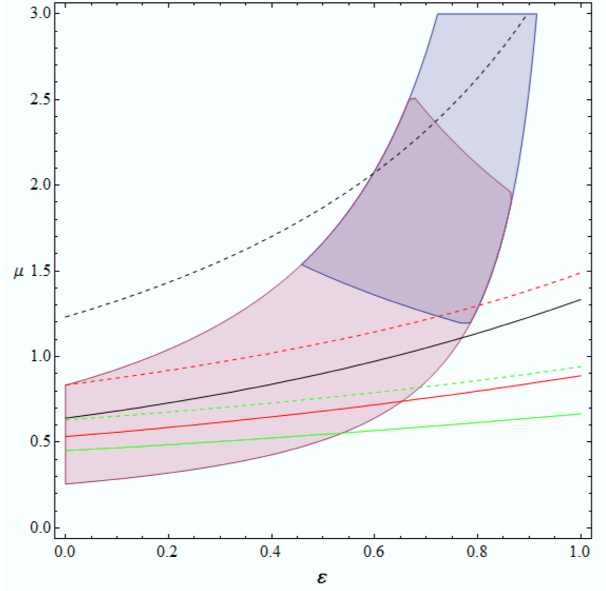}
  \caption{Hyperaccreting ADIOS models in the $\varepsilon-\mu$ plane, superimposed on the combined parameter space for $\gamma = 4/3$ winds and breezes from Fig.~\ref{fig:windbreeze43}. Solid curves show breeze  solutions for (top to bottom) $\dot m_{\rm out} = 0.5, 1, 2$, while dashed curves show wind solutions for the same values of $\dot m_{\rm out}$.  Self-consistent solutions are possible only where a wind or breeze curve crosses the corresponding shaded region.  Thus, winds are possible only for $\dot m_{\rm out}\ltorder 0.75$, whereas breezes are possible for all three values of $\dot m_{\rm out}$.  The type of solution chosen may depend on the outer boundary conditions of the flow.}
  \label{fig:hyperwindbreeze}
\end{figure}

To put this on a more quantitative footing, we consider solutions for $\gamma = 4/3$, $\eta = 0$ (as seems reasonable for winds driven primarily by the effects of radiation pressure).   Figure \ref{fig:hyperwindbreeze} shows solution curves for eq.~(\ref{Rout}) in the $\varepsilon-\mu$ plane for different values of $\dot m_{\rm out}$, superimposed on the combined parameter space for winds and breezes from Fig.~\ref{fig:windbreeze43}.  All three breeze solutions (solid curves) are self-consistent where they cross the lower shaded (breeze) region.  Wind solutions with $\dot m_{\rm out} \ltorder 0.75$ can also be self-consistent, since these cross the two upper shaded regions where winds are permitted.  However, according to our simple model there are no wind solutions for $\dot m_{\rm out} = 1$.
 
A striking feature of the wind curves is their sensitivity to the value of $\dot m_{\rm out}$.  As noted earlier, we generally expect $\dot m_{\rm out} \sim O(1)$, since this is the condition that radiation trapping be marginally effective at the outer edge of the ADIOS.  But it makes a qualitative difference whether the value of $\dot m_{\rm out}$ is 1/2 or 1 --- the former case permits either breezes or winds, whereas only breezes seem to be compatible with the latter.  Our results suggest that there may be two distinct modes of hyperaccretion --- breezes or winds --- and opens the possibility of a system switching between them.  We will investigate the implications of this possible bimodal behavior in a separate paper.

\section{Discussion and conclusions}

We have presented a new version of the adiabatic inflow-outflow (ADIOS) model for radiatively inefficient accretion, treating the outflow on a similar footing to the inflow.  Instead of matching a laminar, adiabatic wind solution to a single-zone model for the inflowing region, we treat the outflow as a ``well-mixed" region characterized by vertically integrated fluxes of mass, energy and angular momentum.  This two-zone formulation is simpler than the original ADIOS models presented in BB99 and BB04, and reflects the global conservation laws in a more transparent way.  More importantly, it avoids the large velocity shears and entropy gradients predicted as a function of height in the original ADIOS model.  If the outflows are highly turbulent, then we argue that a vertically averaged description of the outflow may be more realistic.

This simple modification of the ADIOS model has fundamental implications for the parameter space of physically realizable systems.  For perfectly adiabatic (nonradiating) systems, it implies that the dependence of the mass flow on radius is always described by $n=1$, i.e., $\dot M \propto R^n = R$, in contrast with the original formulation in which $n$ is an undetermined parameter with a value between 0 and 1.  Moreover, a two-zone ADIOS model with a well-mixed outflow requires a conserved energy flux to propagate through {\it both} the inflowing and outflowing zones.  Absent this external energy supply, there is not enough power to propel the outflow, while the inflowing gas is unable to relinquish enough energy to become bound (in the sense that its Bernoulli function is negative).  In the original ADIOS model, where this energy flux is set to zero, the inflow is able to remain bound only because it interacts with the local streamlines of the outflow, which carry much less energy per unit mass than the streamlines originating much closer to the central object. But in the two-zone model, large amounts of mass injected into the outflow at large $R$ mix with large quantities of energy liberated much closer in.

We identify the conserved energy flux with the energy released by the flux of matter ($\dot M_0$) that actually reaches the black hole.  We characterize this energy flux by two parameters, an overall energy efficiency $\mu$ defined by the ratio of the energy flux to ${\rm G}M\dot M/R$ (which is independent of radius), and a parameter $\varepsilon$, describing the fraction of $\mu$ that propagates through the outflow.  Physical models exist only for a fixed range within the interval $0< \varepsilon<1$ that narrows with increasing $\mu$.  If $\mu$ could reach values $\gg 1$, the required value of $\varepsilon$ would approach 1, implying that nearly all the energy must propagate through the outflow.  We stress that $\mu$ is not an accretion efficiency in the usual sense, because it compares the energy liberated by the accreted flux $\dot M_0$ to the binding energy of distant matter that never reaches the black hole.  If the actual accretion efficiency is $\epsilon$, so that the energy flux is $F_E = \epsilon \dot M_0 c^2 / (1-\epsilon)$, then the accretion rate is 
\be 
\label{mdotzero}
\dot M_0  = {1-\epsilon \over \epsilon} \mu {{\rm G}M\dot M \over c^2 R} .  
\ee
A larger value of $\mu$ merely means that equality between $\dot M \propto R$ and $\dot M_0=$ const.~occurs farther from the black hole.  In this case, there would be a wide radial zone in which $\dot M_0$ dominates the mass flow.  Such a region would resemble a unidirectional accretion flow (similar to the early advection-dominated accretion flow [ADAF] models of Narayan \& Yu [1994, 1995]), rather than an inflow-outflow system, and would therefore develop a positive Bernoulli function (BB99).  For this reason, we expect that two-zone ADIOS models are restricted to values of $\mu \ltorder 2$ or 3.    
  
Because our model is self-similar, we cannot treat boundary conditions in detail.  However, we suspect that the value of $\mu$ (and therefore the accretion rate into the black hole) and possibly the value of $\varepsilon$ can adjust in response to the energetic demands of the system.  A couple of our results suggest this.  First, when weak radiative losses are introduced into the model in a self-similar way, the mass flux index $n$ drops below 1 by an amount that depends on the fractional energy loss per decade of radius.  The effect of this is that $\dot M_0$ determines the energy flux through the flow at small radii, but that the flow gradually ``forgets" about the inner boundary with increasing $R$, i.e., as the centrally injected energy is radiated away.  The case of hypercritical accretion provides a more dramatic example, since solutions with a large ratio of outer-to-inner radius must have $\dot m_a \sim 1 + \ln (R_{\rm out} /R_{\rm in})$.  Apparently, matter supplied to a black hole at a highly super-Eddington rate will also be accreted at a highly super-Eddington rate, in order to supply enough energy to match radiative losses in addition to powering the flow.  Nevertheless, both $\mu$ and $\varepsilon$ must fall into fairly narrow ranges in order for the flows to be self-consistent. (The parameter $\beta$, which describes energy transfer from inflow to outflow, must also self-regulate to $\approx 1$ for $n=1$). 

Despite the highly constrained parameter space, our solutions display a striking range of behaviors from gentle breezes, governed by viscous stress, to transsonic winds.  Such bimodal behavior extends to hypercritical accretion, where the dependence of radiative losses on flow geometry and dynamics breaks self-similarity close to the trapping radius.  We note that Shakura \& Sunyaev (1973) foresaw the possibility of both powerful winds and gentle outflows in their pioneering discussion of hypercritical accretion. 

Both breezes and winds are unbound, in the sense that the Bernoulli function is positive, but any significant radiative losses near the outer radii of the ADIOS could quench the weaker outflows.  It is possible that breeze solutions represent large-scale circulations rather than genuine mass loss, in which case the total mass of the system would grow with time if matter continues to be introduced from outside.  
 
The strength of the ADIOS approach is that it relies on simple conservation laws and dynamical equilibrium conditions, which impose clear constraints on the space of possible flows. Thus it can provide a useful framework against which numerical models or observations can be analyzed.  Beyond the application of conservation laws and dynamical relations, the ADIOS idea relies on the assumption that some physical mechanism exists to bifurcate the available energy and to launch an outflow that is spatially distinct from the inflowing gas.  In the case of the two-zone models presented here, it also assumes that the inflow and outflow zones are both well-mixed.  These assumptions may not be valid, or may apply only in certain types of systems (e.g., depending on boundary conditions, magnetic field structures [see, e.g., Beckwith, Hawley \& Krolik 2008, 2009], or the relative importance of magnetic and radiative stresses).  But the two-zone conservation laws and dynamical relations are very general and can be adapted to treat alternative assumptions about the flow pattern.  For example, one can devise a two-fluid model that might capture the physics of a radiatively inefficient flow dominated by nonlinear convection, with the inflow and outflow zones interpenetrating one another.  Such a model could represent a ``convection-dominated accretion flow" (CDAF: Narayan, Igumenshchev \& Abramowicz 2000; Quataert \& Gruzinov 2000).  We will present such a model, and compare its properties to those of the two-zone ADIOS model, in a forthcoming paper.     

\section*{Acknowledgements}
I thank Phil Armitage, Kris Beckwith and Roger Blandford for helpful discussions. This work was supported in part by NSF grant AST-0907872.

\bsp
\end{document}